\documentclass[twocolumn,prb,superscriptaddress]{revtex4}
\usepackage{graphicx}
\usepackage{dcolumn}
\usepackage{bm}
\usepackage{amsmath}
\usepackage{times}
\usepackage{epstopdf}
\usepackage{color}
\usepackage[breaklinks=true,colorlinks,citecolor=blue,linkcolor=blue,urlcolor=blue]{hyperref}
\makeatletter

\newcommand{\Rmnum}[1]{\expandafter\@slowromancap\romannumeral #1@}
\makeatother
\graphicspath{{images/}}
\begin{document}
\author{Mahammad Tahir}
\affiliation{Department of Physics, Indian Institute of Technology Kanpur, Kanpur- 208016, India}
\author{Swati Pandey}
\affiliation{Department of Physics, Indian Institute of Technology Kanpur, Kanpur- 208016, India}
\author{Sourabh Manna}
\affiliation{Natural Sciences and Science Education, National Institute of Education, Nanyang Technological University, Singapore-637616 Singapore}
\author{Rajdeep Singh Rawat}
\affiliation{Natural Sciences and Science Education, National Institute of Education, Nanyang Technological University, Singapore-637616 Singapore}
\author{Rohit Medwal} 
\email{rmedwal@iitk.ac.in}
\affiliation{Department of Physics, Indian Institute of Technology Kanpur, Kanpur- 208016, India}
\author{Soumik Mukhopadhyay}
\email{soumikm@iitk.ac.in}
\affiliation{Department of Physics, Indian Institute of Technology Kanpur, Kanpur- 208016, India}
\title{Probing Interfacial Spin Dynamics and Temperature Dependent Asymmetry in Spin Pumping Across $\mathrm{Ni_{80}Fe_{20}/Cu/Cr_{1.12}Te_{2}}$ Interfaces}
\begin{abstract}

Spin transfers in magnetic multilayers offers a promising pathway toward ultrafast, energy efficient spintronic devices. In this study, we investigate the interfacial spin pumping and temperature dependent spin current exchange in a $\mathrm{Cr_{1.12}Te_{2}/Cu/Ni_{80}Fe_{20}}$ (Py)(FM1/NM/FM2) trilayer structure. Using broadband and cryogenic ferromagnetic resonance (FMR) measurements, we investigate key magnetization dynamical parameters, including the effective Gilbert damping factor, effective magnetic fields, interfacial spin mixing conductance, and spin current density. Efficient spin angular momentum transfers from Py to $\mathrm{Cr_{1.12}Te_2}$ are observed at room temperature. At lower temperatures, the enhanced linewidth reflects temperature dependent spin pumping effects occurring at distinct precession frequencies of the ferromagnetic layers. Notably, the absence of interfacial damping indicates that spin pumping can be modulated by controlling the net spin current flow. These findings offer critical insight into temperature dependent tunable spin transport mechanisms in magnetic multilayers, highlighting their potential for next  generation spintronic applications.
\end{abstract}

\maketitle
\section{Introduction}
Pure spin current generation, transportation, and detection represent core mechanisms in spintronics, enabling the realization of high-efficiency spintronic memory and computational devices~\cite{Urazhdin5,Locatelli5,Sankey5}. Among the various methods, spin pumping is a prominent mechanism wherein a ferromagnet (FM), driven into ferromagnetic resonance (FMR), emits a pure spin current into adjacent layers~\cite{Brataas5}. This charge-free approach generates spin transfer torque (STT) as the spin current traverses a normal metal (NM) spacer, either reflecting back to the source FM or interacting with a second FM layer to modulate its magnetization dynamics~\cite{Stiles5}. Spin pumping leads to an enhancement of the effective Gilbert damping due to the angular momentum loss from the precessing FM, typically observed as a broadening of the FMR linewidth~\cite{Mizukami5,Tserkovnyak5.1,Tserkovnyak5.2,Tserkovnyak5.3}. Detection of spin currents is often realized through the inverse spin Hall effect (ISHE) in a spin sink layer, or via element-specific techniques such as x-ray magnetic circular dichroism (XMCD), which probes spin accumulation and magnetization precession~\cite{Marcham5, Baker5.1, Kukreja5}. Unlike mechanisms like nonlocal spin injection~\cite{Kimura5} or the spin Hall effect~\cite{Valenzuela5}, which require charge current injection, spin pumping operates purely via non-equilibrium magnetization dynamics, enabling spin current generation irrespective of the electronic conduction properties of the materials~\cite{Wei5,HWang5,Gupta5}.

In a $\mathrm{FM1/NM/FM2}$ trilayer system with collinearly aligned magnetizations, the transverse spin current emitted by the source layer (FM1) is absorbed by the sink layer (FM2), exerting a spin transfer torque (STT) that contributes to the enhanced effective Gilbert damping in FM1. Such interlayer spin exchange mechanisms are vital for the operation of magnetic read head devices~\cite{Tsang5} and spin transfer torque nano oscillators (STNOs)~\cite{Silva5}. Experimental studies on $\mathrm{FM1/NM/FM2}$ structures have revealed a range of complex spin transport behaviors, including long-range dynamic spin coupling~\cite{Bauer5.1}, bias and alignment-dependent damping modulation~\cite{Salikhov5.1}, and domain wall-mediated spin interactions~\cite{Salikhov5.2}. These phenomena are further influenced by anisotropic spin relaxation under varied geometric configurations~\cite{Bailey5}, direction-dependent spin current absorption~\cite{Baker5.2}, and asymmetric spin pumping effects in trilayer systems~\cite{Yang5,Rohit5}. Collectively, these findings highlight the intricate and highly tunable nature of pure spin current dynamics in multilayered magnetic heterostructures~\cite{Yang5}.
\begin{figure*}
\includegraphics[width=0.99\linewidth]{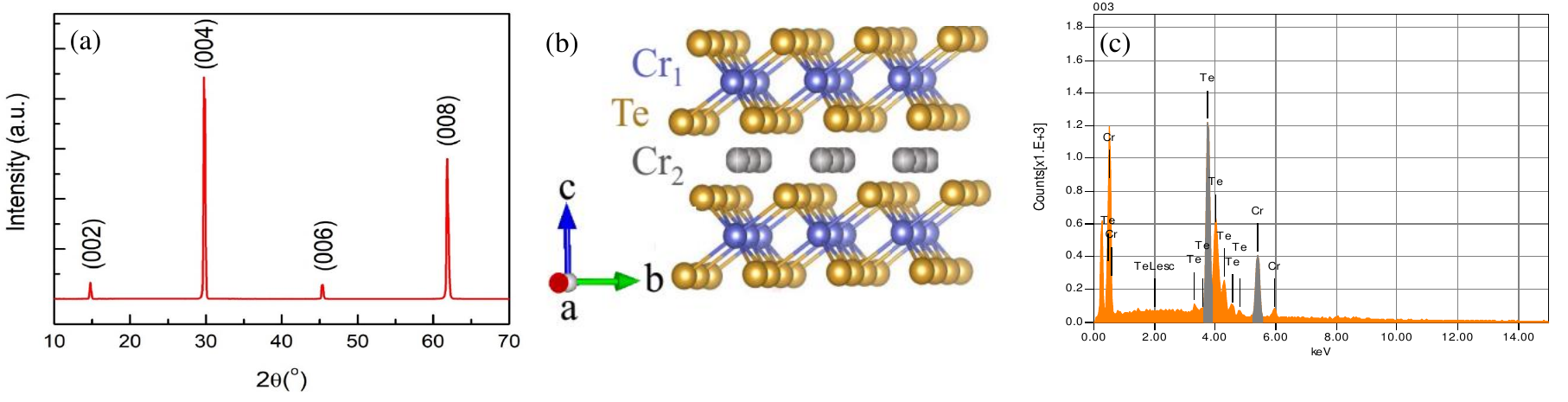}
\caption{(a) XRD pattern of $\mathrm{Cr_{1.12}Te_{2}}$. (b) and crystal structure of $\mathrm{Cr_{1.12}Te_{2}}$ (c) EDX spectrum of single crystal.}
\label{fig1}
\end{figure*}
\begin{figure}
\includegraphics[width=0.99\linewidth]{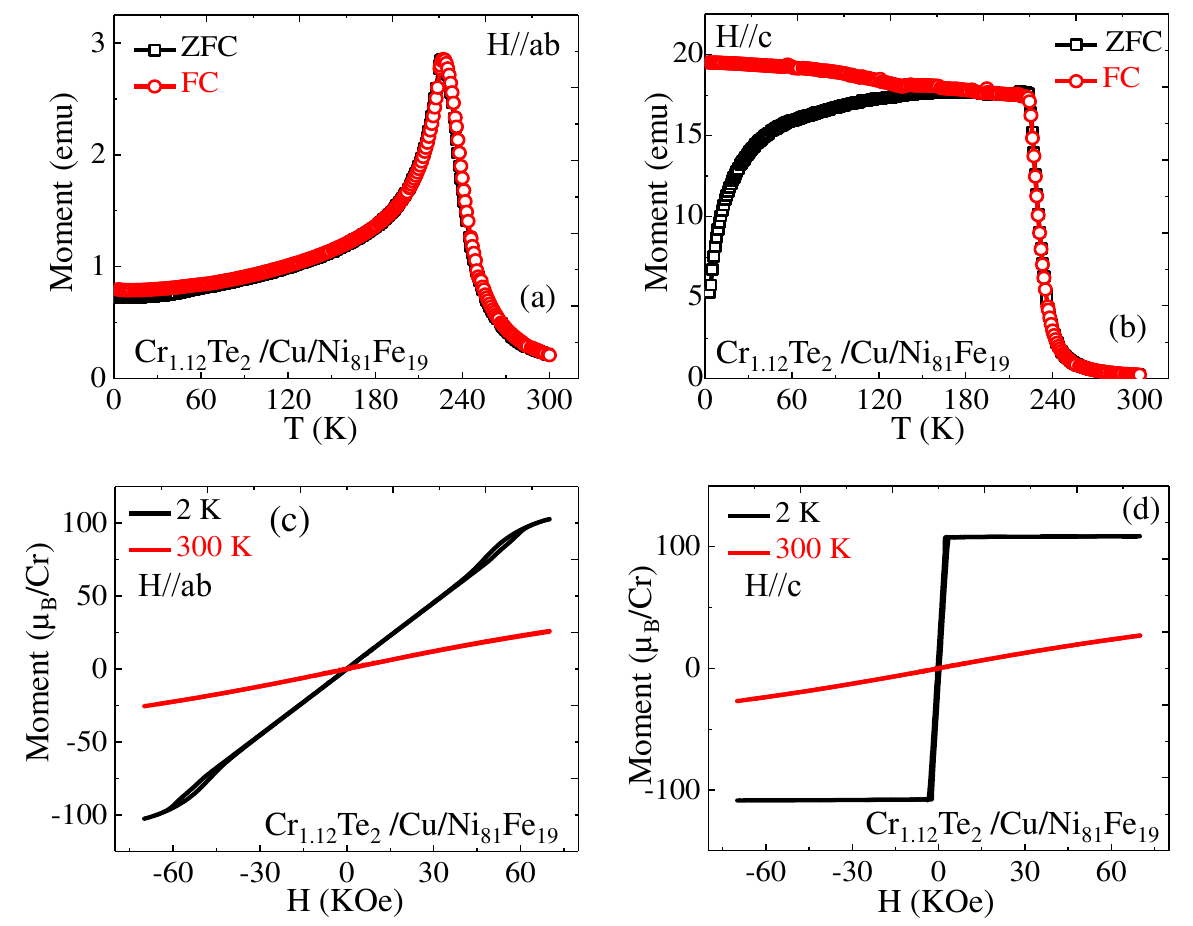}
\caption{(a) Shows $\chi(T)$ along (a) $H \parallel ab$ (b) $H \parallel c$. M(H) curves along (c)  $H \parallel ab$ and (d) $H \parallel c$ measured at 2 K and 300 K. }
\label{fig2}
\end{figure}

Two-dimensional (2D) magnetic van der Waals (vdW) materials have recently emerged as a critical class of quantum materials in condensed matter physics, owing to their intrinsic spin-dependent interactions and suitability for next-generation spintronic and optoelectronic applications~\cite{Novoselov5,Huang5,Cortie5,Hu5,Kurebayashi5}. Considerable progress has been achieved in the rational design and comprehensive characterization of structural, magnetic, and spectroscopic investigations of novel vdW magnets, including $\mathrm{CrXTe_{3}}$ (X = Si, Ge, Sn)~\cite{Sivadas5,Zhuang5,Casto5}, $\mathrm{Fe_{x}GeTe_{2}}$ (x = 3, 4, 5) ~\cite{May5,Mondal5}, chromium trihalides such as $\mathrm{CrI_{3}}$~\cite{Besbes5,Xu5}, nonstoichiometric chromium tellurides $\mathrm{Cr_{1+x}Te_{2}}$ ~\cite{Fujisawa5, Wang5.1, Jiang5, Liu5, Wang5.2, Zhao5, Cao5, Liu5.1, Liu5.2, Zhang5.1, Mohamed5, Shu5, Wang5.3, Saha5}. These systems are of particular interest due to their low-dimensional magnetism, tunable spin configurations, and relevance to quantum spintronic frameworks~\cite{Fujisawa5, Wang5.1, Jiang5, Liu5, Wang5.2, Zhao5, Cao5, Liu5.1, Liu5.2, Zhang5.1, Mohamed5, Shu5, Wang5.3, Saha5}. Among them, $\mathrm{Cr_{1+x}Te_{2}}$ has garnered significant attention for its robust ferromagnetism with Curie temperatures ($\mathrm{T_{C}}$) surpassing ambient conditions in compositions such as $\mathrm{Cr_{1.75}Te_{2}}$~\cite{Liu5.2}, alongside distinctive magnetic and transport phenomena~\cite{Saha5}. Notably, the structural and magnetic behavior of $\mathrm{Cr_{1+x}Te_{2}}$ is markedly sensitive to compositional stoichiometry, critically influencing phase stability and magnetic ordering~\cite{Bensch5}. In the $\mathrm{Cr_{1+x}Te_{2}}$ series, excess Cr atoms (where $x > 0$) are intercalated within the van der Waals gaps along the crystallographic c-axis, occupying sites between adjacent stoichiometrically complete $\mathrm{CrTe_{2}}$ bilayers. At elevated Cr concentrations  ($x \geq 0.5$), these materials show enhanced $\mathrm{T_{C}}$ and exhibit in-plane magnetic anisotropy, with the easy axis aligned along the ab-plane. Such compositions have garnered considerable attention due to the emergence of rich spintronic phenomena, including the anomalous Hall effect~\cite{Purwar5}, the topological Hall effect~\cite{Purwar5}, and tunable interlayer magnetic coupling~\cite{Zhang5.2, Wen5, Coughlin5}. Conversely, as Cr intercalation is reduced, a notable suppression of ($\mathrm{T_{C}}$) to 160 K is observed, accompanied by a spin reorientation transition wherein the easy axis realigns along the c-axis~\cite{Jiang5}. Compounds in this family, including CrTe (x = 1), $\mathrm{Cr_{2}Te_{3}}$ (x = 0.33),$\mathrm{Cr_{3}Te_{4}}$ (x = 0.5), $\mathrm{Cr_{5}Te_{8}}$ (x = 0.25), and $\mathrm{CrTe_{2}}$ (x = 0), exhibit layered structures comprised of alternating Cr and Te atomic planes. The Cr occupancy within every second Cr layer diminishes systematically with decreasing $x$, directly influencing magnetic ordering temperatures~\cite{Lasek5}. Reported $\mathrm{T_{C}}$ include 340 K for $\mathrm{Cr_{3}Te_{4}}$, 240 K for $\mathrm{Cr_{5}Te_{8}}$, and 195 K for $\mathrm{Cr_{2}Te_{3}}$~\cite{Lasek5}. This tunability originates from compositional modulation of the lattice parameter $c$, which governs magnetic anisotropy energy and exchange interactions~\cite{Hak5}. Despite extensive studies on compositions with moderate to high Cr content, the low Cr regime $x \leq 0.2$ remains relatively unexplored. In particular $\mathrm{CrTe_{2}}$, (x = 0), which lacks Cr intercalation entirely, has received limited attention, primarily due to significant synthetic challenges associated with stabilizing this phase under ambient or conventional growth conditions.

In this study, we systematically examine the spin transport characteristics of the newly synthesized 2D magnetic compound $\mathrm{Cr_{1.12}Te_2}$. This material undergoes a transition to a ferromagnetically (FM) ordered state below its Curie temperature $\mathrm{T_{C}}$ = 250 K, whereas it retains a paramagnetic (PM) phase at temperatures exceeding $\mathrm{T_{C}}$. To investigate spin dynamics in this system, we utilize a spin valve like trilayer structure composed of Py/Cu/$\mathrm{Cr_{1.12}Te_2}$, where $\mathrm{Ni_{80}Fe_{20}}$ (Py) serves as the ferromagnetic spin injector (FM1), Cu acts as the NM spacer, and $\mathrm{Cr_{1.12}Te_2}$ functions as both a spin sink (in its PM and FM states) and as a  source of spin current (FM2) in the low-temperature regime. We demonstrate efficient spin current injection from the Py layer into  $\mathrm{Cr_{1.12}Te_2}$, via spin pumping at room temperature, where  $\mathrm{Cr_{1.12}Te_2}$, remains in the PM state. Upon cooling below ($\mathrm{T_{C}}$), the material transitions to an FM state with long-range spin order. This temperature-driven magnetic phase transition enables a systematic study of spin current behavior across both PM and FM regimes. Additionally, we observe that in the FM phase,  $\mathrm{Cr_{1.12}Te_2}$, can also act as a source of spin current, enabling temperature-dependent spin pumping back into the Py layer. By analyzing the temperature dependence of the spin pumping phenomenon, we reveal how the evolving magnetic state of  $\mathrm{Cr_{1.12}Te_2}$, affects spin current absorption and transport. These findings offer valuable insights into the coupling between spin dynamics and magnetic phase transitions in low-dimensional van der Waals magnets.

\begin{figure*}
\includegraphics[width=0.99\linewidth]{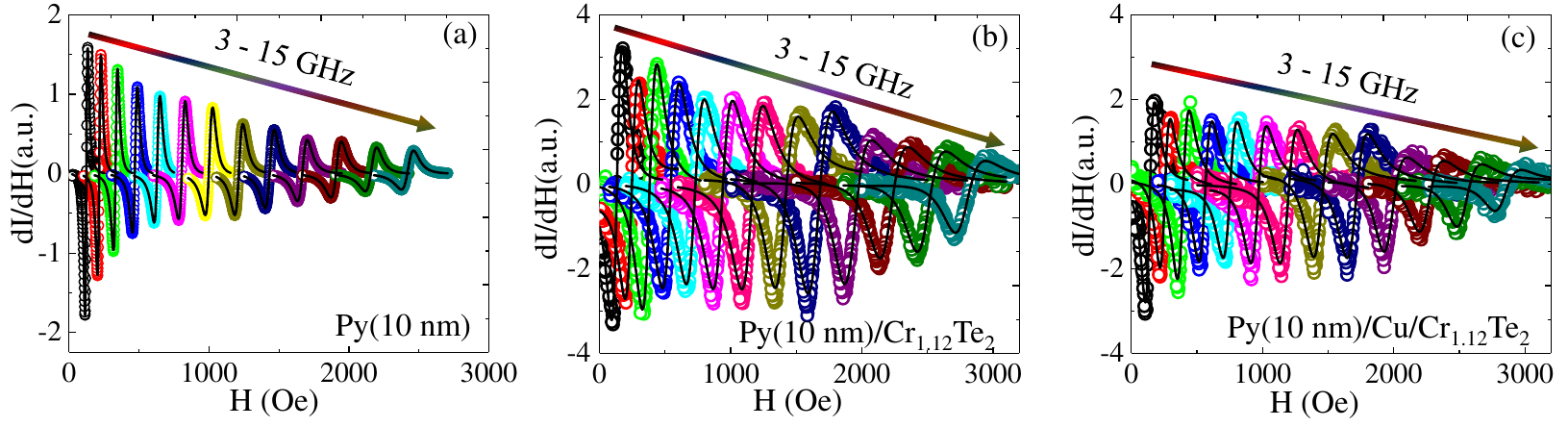}
\caption{(a) FMR spectra recorded on (a) $\mathrm{Si/SiO_{2}}$/Py(10 nm), (b) Py(10 nm)$\mathrm{/Cr_{1.12}Te_{2}}$ and (c) Py(10 nm)/Cu(2 nm)$\mathrm{/Cr_{1.12}Te_{2}}$ samples. Symbols are experimental data points, and the black solid line is the fitting using Eq.~\ref{eq1}.}
\label{fig3}
\end{figure*}
\section{Sample preparation details and characterization}
High quality single crystals $\mathrm{Cr_{1.12}Te_2}$ were synthesized using a chemical vapor transport (CVT) method, with iodine serving as the transport agent. Stoichiometric quantities of high purity chromium (99.99\%) and tellurium (99.999\%) were weighed according to the nominal composition and thoroughly mixed under an inert argon atmosphere to prevent oxidation. The homogenized precursor mixture, along with iodine ($\approx 5 \mathrm{mg/cm}^3$), was sealed under high vacuum in a quartz ampoule to ensure a contamination free environment. The sealed ampoule was placed in a dual zone horizontal tube furnace configured to establish a thermal gradient. The hot zone was heated to $1000^{\circ}$ C and the cold zone was maintained at, $820^{\circ}$ C over a 12 hour ramping period. This temperature gradient promoted directional vapor transport from the hot to the cold zone, facilitating a single crystal growth. The synthesis was sustained for 10 days to ensure the formation of large, well-faceted, clean single crystals. After the growth period, the furnace was allowed to cool to room temperature, and the crystals were retrieved from the cold end of the ampoule for subsequent characterization. Figure~\ref{fig1}a shows the single crystal X-ray diffraction (XRD) pattern of a $\mathrm{Cr_{1.12}Te_2}$ single crystal. The presence of only sharp and intense $\mathrm{00l}$ reflections confirms high crystallinity and strong preferential orientation along the c-axis, indicating that the dominant crystal facet lies parallel to the ab plane. Figure~\ref{fig1}b depicts the layered structure, highlighting excess Cr intercalated within Cr deficient vdWs gaps along the c-axis between alternating $\mathrm{CrTe{_2}}$ layers. EDX analysis, presented in Figure~\ref{fig1}c, reveals a Cr-to-Te atomic ratio of 35.84:64.14, consistent with Cr-rich stoichiometry.

To investigate the interfacial spin dynamics and temperature dependent asymmetric spin current exchange, we prepared the Py(10 nm)/Cu/$\mathrm{Cr_{1.12}Te_2}$,
Py(10 nm)/$\mathrm{Cr_{1.12}Te_2}$, $\mathrm{Si/SiO_2/Py}$(10 nm)/Cu, and $\mathrm{Si/SiO_2/Py}$(10 nm) samples were deposited using DC magnetron sputtering on precleaned Si/SiO$_{2}$ substrates and on $\mathrm{Cr_{1.12}Te_2}$ single crystals with a base pressure better than 3 × 10$^{-7}$ Torr. The deposition pressure and argon flow rate were 3 × 10$^{-3}$ Torr and 15 standard cubic centimeters per minute (SCCM), respectively. High quality single crystals $\mathrm{Cr_{1.12}Te_2}$ have been employed as substrates for thin film deposition and serve multiple functional roles depending on the temperature regime. At room temperature, they act as efficient spin sink materials for the detection of spin pumping phenomena. In contrast, at low temperatures, these crystals exhibit ferromagnetic ordering and are utilized as a ferromagnetic layer in the Py(10 nm)/Cu/$\mathrm{Cr_{1.12}Te_2}$, sample.

Figure~\ref{fig2} (a) shows the magnetic susceptibility, $\chi(T)[ = M/H]$, measured during zero field cooling (ZFC) and field cooling (FC) under an applied magnetic field of 0.01 T along the ab plane for the Py/Cu$\mathrm{/Cr_{1.12}Te_2}$ sample. Figure~\ref{fig2}(b) presents the corresponding susceptibility measurements for the same sample with the magnetic field applied along the c axis. Upon lowering the temperature, anisotropic behavior in magnetization emerges below 250 K, as evidenced by the divergence in magnetic susceptibility, $\chi(T)$, between the two orientations: $H \parallel c$ and $H \parallel ab$. A clear FM transition is observed at $\mathrm{T_C} \approx 250$ K for $H \parallel c$, while for $H \parallel ab$, the transition occurs at a slightly higher temperature of approximately 270 K. Figure~\ref{fig2}(c),(d) presents the isothermal magnetization curves, M(H), measured at 2 K and 300 K for the 
Py/Cu$\mathrm{/Cr_{1.12}Te_2}$ sample. The data show that magnetic saturation is achieved at lower applied fields for 
$H \parallel c$ direction compared to $H \parallel ab$ , indicating that the c-axis is the magnetic easy axis. This anisotropic behavior is attributed to a lower magnetic anisotropy energy along the c-axis, favoring the alignment of magnetic moments in this direction.

\section{Results and discussion}

\section{Observation of spin pumping at $\mathrm{Py/Cu/Cr_{1.12}Te_{2}}$ interface at room temperature}

The interfacial spin dynamics of all samples were investigated at room temperature using the flip-chip ferromagnetic resonance (FMR) technique. FMR spectra recorded at room temperature for three different samples, Py(10 nm), Py(10 nm)$\mathrm{/Cr_{1.12}Te_{2}}$ and  Py(10 nm)/Cu(2 nm)$\mathrm{/Cr_{1.12}Te_{2}}$ are shown in Fig.~\ref{fig3}a–c. The FMR spectra were fitted using a derivative Lorentzian function incorporating both symmetric and antisymmetric components, as described by Eq.~\ref{eq1}~\cite{Woltersdorf5}. The experimental data are depicted by symbols, while the black solid lines represent fits based on Eq.~\ref{eq1}. 

\begin{widetext}
\begin{equation}
\frac{d \mathrm{I}_{\mathrm{FMR}}}{d \mathrm{H}} = 4\mathrm{A}\frac{\Delta \mathrm{H}( \mathrm{H}- \mathrm{H}_{\mathrm{res}})}{(4( \mathrm{H}- \mathrm{H}_{\mathrm{res}})^2+
(\Delta  \mathrm{H})^2)^2}-\mathrm{S}\frac{(\Delta  \mathrm{H})^2-4(\mathrm{H}-\mathrm{H}_{\mathrm{res}})^2}{(4( \mathrm{H}- \mathrm{H}_{\mathrm{res}})^2+(\Delta  \mathrm{H})^2)^2},\label{eq1}
\end{equation}
\end{widetext}

\begin{figure}
\includegraphics[width=0.99\linewidth]{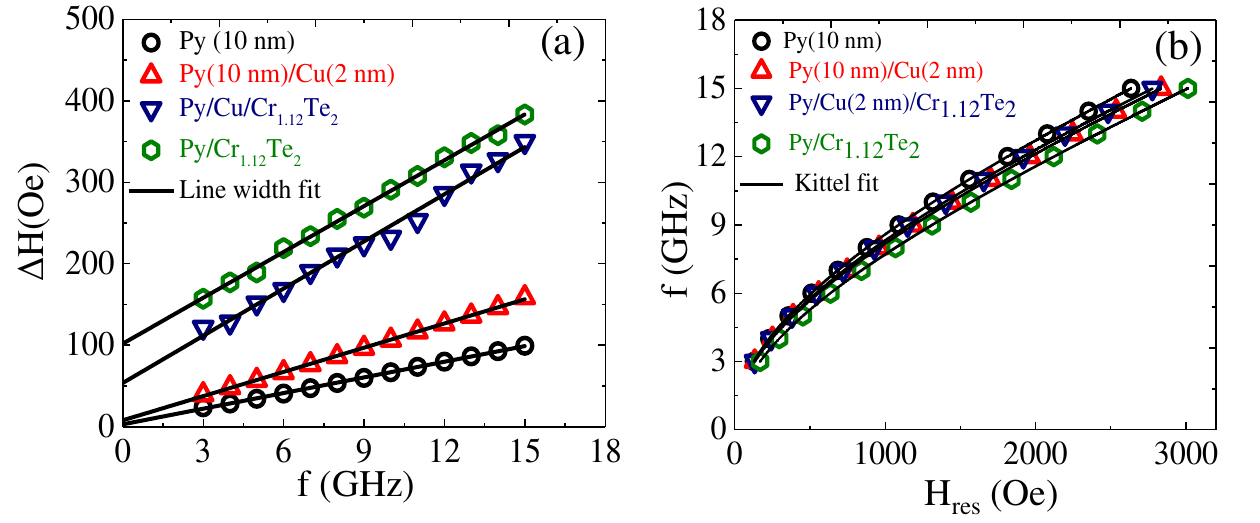}
\caption{(a) Linewidth ($\Delta \mathrm{H}$) vs. frequency ($\mathrm{f}$) for Py(10 nm), Py(10 nm)/Cu(2 nm), Py(10 nm)$\mathrm{/Cr_{1.12}Te_{2}}$ and Py(10 nm)/Cu(2 )/$\mathrm{Cr_{1.12}Te_{2}}$ samples. The data points are extracted from FMR fittings, and the solid lines are the fittings corresponding to the linewidth equation (Eq.~\ref{eq2}). (b) Resonance field $\mathrm{H}_{\mathrm{res}}$ vs. frequency $\mathrm{f}$ for all the samples. Black solid lines are the fits using Eq.~\ref{eq3}.}
\label{fig4}
\end{figure}

Here, $\mathrm{H}$ referring to the in-plane applied DC magnetic field, $\Delta \mathrm{H}$ is the full width at half maximum (FWHM) of the FMR signal is known as the linewidth and $\mathrm{H}_{\mathrm{res}}$ is the resonance field at which maximum microwave absorption occurs. The amplitudes $\mathrm{S}$ and $\mathrm{A}$ of the FMR signal correspond to the symmetric and antisymmetric components, respectively~\cite{Woltersdorf5}. The values of $\Delta \mathrm{H}$ and $\mathrm{H}_{\mathrm{res}}$ were extracted from fits to the FMR spectra. The FMR linewidth may be broadened by several factors, including surface and interfacial roughness, as well as structural defects within the FM thin films. Such imperfections contribute to enhanced magnetic relaxation~\cite{Kittel5, Gilbert5}. The total linewidth  $\Delta \mathrm{H}$  can be expressed as:
\begin{equation*}
\Delta \mathrm{H} = \Delta \mathrm{H}_{\mathrm{Intrinsic}} + \Delta \mathrm{H}_{\mathrm{Extrinsic}},
\end{equation*}
The intrinsic contribution to the FMR linewidth originates from effective Gilbert damping, which is governed by fundamental energy dissipation processes in the magnetization dynamics. This mechanism is described by the Landau Lifshitz Gilbert (LLG) equation. According to this framework, the FMR linewidth exhibits a linear dependence on the microwave absorption frequency. The frequency f dependence of the linewidth $\Delta \mathrm{H}$ is presented in Fig.~\ref{fig4}a, where the solid line represents a linear fit based on the following relation~\cite{Kittel5, Gilbert5}:

\begin{eqnarray}
\Delta \mathrm{H} = \frac{4\pi \alpha_{\mathrm{eff}}}{\gamma} \mathrm{f}+\Delta\mathrm{H}_0,\label{eq2}
\end{eqnarray}

The term $\Delta\mathrm{H}_0$ represents the frequency-independent contribution to the linewidth, attributed to inhomogeneous linewidth broadening arising from magnetic nonuniformities within the thin film. The observed linear dependence of $\Delta \mathrm{H}$ on frequency $\mathrm{f}$, as illustrated in Fig.~\ref{fig4}a, shows that the effective Gilbert damping in our Py/Cu$\mathrm{/Cr_{1.12}Te_2}$, Py$\mathrm{/Cr_{1.12}Te_2}$, Py/Cu, and ref. Py samples predominantly originates from intrinsic mechanisms. The effective Gilbert damping constant, $\alpha_{\mathrm{eff}}$, determined from the slope of the linear dependence of $\Delta \mathrm{H}$ on $\mathrm{f}$ has the following contributions:

\begin{equation*}
\alpha_{\mathrm{eff}}=\alpha_{\mathrm{intrinsic}}+\alpha_{\mathrm{extrinsic}},
\end{equation*}

Here, $\alpha _{\mathrm{intrinsic}}$ denotes the damping component associated with energy dissipation into the lattice within the bulk of the Py layer~\cite{Hickey5}. In contrast, the extrinsic damping parameter, $\alpha_{\mathrm{extrinsic}}$, is primarily attributed to spin pumping mechanisms, wherein spin angular momentum is transferred via spin current emission from the Py layer into the adjacent $\mathrm{Cr_{1.12}Te_2}$ and $\mathrm{Cu/Cr_{1.12}Te_2}$ interfaces. This process is further influenced by interfacial spin-flip scattering mediated by spin–orbit coupling at the interface~\cite{Conca5}. The introduction of a Cu interlayer is a well-established strategy in spintronic devices, employed to tailor interfacial resistance and enhance spin current injection by modulating spin transmission characteristics. In the present study, the Cu interlayer serves to decouple the $\mathrm{Cr_{1.12}Te_2}$ from the Py layer, thereby mitigating direct interfacial exchange interactions. Furthermore, the inclusion of Cu helps suppress proximity-induced ferromagnetism in the adjacent layer during spin pumping measurements. The thickness of the Cu layer in our samples is kept below its known spin diffusion length, thereby ensuring efficient spin transport across the interface. As shown in Fig.~\ref{fig4}a, the frequency dependence of the FMR linewidth $\Delta \mathrm{H}$ vs $\mathrm{f}$ is comparable for both the Py$\mathrm{/Cr_{1.12}Te_2}$ and Py/Cu$\mathrm{/Cr_{1.12}Te_2}$ interfaces, indicating that the insertion of the Cu layer does not affect bulk related $\alpha _{\mathrm{intrinsic}}$. Figure~\ref{fig4}b shows the variation of the resonance field $\mathrm{H}_{\mathrm{res}}$ as a function of the microwave excitation frequency $\mathrm{f}$. The effective magnetization $4\pi \mathrm{M}_{\mathrm{eff}}$ and $\mathrm{H}_{\mathrm{K}}$ the in-plane anisotropy field of the Py layer are quantitatively determined by fitting the experimental data using the following Kittel equation (Eq.~\ref{eq3})~\cite{Kittel5}.

\begin{equation}
\mathrm{f} = \frac{\gamma}{2\pi}\left[\left( \mathrm{H}_{\mathrm{res}}+ \mathrm{H}_{\mathrm{k}}\right) \left( \mathrm{H}_{\mathrm{res}}+ \mathrm{H}_{\mathrm{k}}+ 4 \pi \mathrm{M}_{\mathrm{eff}}\right)\right]^{1/2},\label{eq3}
\end{equation}
Here $\gamma$ = $\frac{g\mu_{B}}{\hbar}$ is the gyromagnetic ratio, $\mu_{B}$ is the Bohr magneton, $g$ is the Lande's spectroscopic splitting factor, and $\hbar$ is the reduced Planck's constant. The values of $4\pi\mathrm{M}_{\mathrm{eff}}$ for all samples are shown in Table~\ref{table5.1}. This $4 \pi \mathrm{M}_{\mathrm{eff}}$ is given by:
\begin{equation*}
4 \pi \mathrm{M}_{\mathrm{eff}} = 4\pi\mathrm{M}_{\mathrm{s}} - \frac{\mathrm{K}_{\mathrm{s}}}{\mathrm{M}_{\mathrm{s}} \mathrm{t}_{\mathrm{Py}}},
\end{equation*}
\begin{table*}
    \centering
        \begin{tabular}{cccccc}
        \hline
        \hline
        Sample & $\gamma$  & $\alpha_{eff}$ & $\Delta\mathrm{H_{0}}$ &
$4\pi\mathrm{M}_{\mathrm{eff}}$ (Oe)&$\mathrm{H_K}$ (Oe) \\
        \hline
Py(10 nm)  & 1.70×$10^2$
 &0.0086$\pm$0.0006 &2.82$\pm$0.49 & 8925.98$\pm$13.84&7.14$\pm$0.10\\
Py(10 nm)/Cu(2 nm) & 1.73×$10^2$
 & $0.0136 \pm 0.0009$ & $7.73 \pm 0.74$ & 8852.88 $\pm$ 13.84 & $0.42 \pm 0.11$\\
Py(10 nm)/$\mathrm{Cr_{1.12}Te_{2}}$ & 1.67×$10^2$
& $0.0246\pm0.0007$ & $53.60 \pm 5.58$ &$8175.42\pm181.82$ & $17.41\pm1.86$\\
Py(10 nm)/Cu(2 nm)/$\mathrm{Cr_{1.12}Te_{2}}$ & 1.65×$10^2$ &
$0.0258\pm0.0001$ & $101.84\pm2.61$ & $8875.65\pm226.04$ & $11.49 \pm 1.92$\\
         \\
        \hline
        \hline
        \end{tabular}
    \caption{Extracted parameters: $\gamma$ (gyromagnetic ratio), $\alpha_{\mathrm{eff}}$ (effective Gilbert damping), $\Delta \mathrm{H_0}$ (inhomogeneous linewidth broadening), $4\pi\mathrm{M}_{\mathrm{eff}}$ (effective magnetization), and $\mathrm{H_K}$ (in-plane anisotropy).}
    \label{table5.1}
\end{table*}
Here, $\mathrm{K}_{\mathrm{s}}$ represents the surface or interface anisotropy constant, $\mathrm{t_{Py}}$ is the thickness of the Py layer, and $\mathrm{M}_{\mathrm{s}}$ denotes the saturation magnetization. The reduction of $4\pi\mathrm{M}_{\mathrm{eff}}$ observed in the Py/Cu$\mathrm{/Cr_{1.12}Te_2}$ and Py$\mathrm{/Cr_{1.12}Te_2}$ interfaces as compared to the ref. Py layer results from the combined influence of interfacial SOC and magnetic interactions at the interface. These $4\pi\mathrm{M}_{\mathrm{eff}}$ values are governed not only by the intrinsic magnetic properties of the Py layer but also by extrinsic effects arising from the adjacent Cu$\mathrm{/Cr_{1.12}Te_2}$ and $\mathrm{Cr_{1.12}Te_2}$ interfaces. The insertion of a thin Cu layer (2 nm) at the Py$\mathrm{/Cr_{1.12}Te_2}$ interface separates both layers and modifies both the interfacial and in plane $\mathrm{H}_{\mathrm{K}}$ anisotropies.

The experimentally observed values of $\alpha_{\mathrm{eff}}$ are measured for Py(10 nm), Py(10 nm)/Cu(2 nm), Py(10 nm)$\mathrm{/Cr_{1.12}Te_{2}}$ and  Py(10 nm)/Cu(2 nm)$\mathrm{/Cr_{1.12}Te_{2}}$ samples and are shown in Table~\ref{table5.1}. A large increase in $\alpha _{\mathrm{eff}}$ is observed for the Py$\mathrm{/Cr_{1.12}Te_{2}}$ and Py/Cu(2 nm)$\mathrm{/Cr_{1.12}Te_{2}}$ interfaces compared to the ref. Py sample, as shown in Table~\ref{table5.1}. The enhancement in the effective Gilbert damping is attributed to spin pumping and quantitatively estimated by $\Delta\alpha = \alpha_{\mathrm{Cr_{1.12}Te_{2}/Cu/Py}} - \alpha_{\mathrm{Py}}$, which is used to calculate the interfacial spin mixing conductance ($\mathrm{g}^{\uparrow\downarrow}$). The interfacial spin mixing conductance ($\mathrm{g}^{\uparrow\downarrow}$), quantifies the efficiency of spin current injection by the precessing magnetization of the Py layer across the Cu (2 nm)$\mathrm{/Cr_{1.12}Te_{2}}$ interface or the $\mathrm{Cr_{1.12}Te_{2}}$  interface and is given by the following expression~\cite{Tserkovnyak5.1,Tserkovnyak5.2,Tserkovnyak5.3}:
\begin{eqnarray}
\mathrm{g}^{\uparrow\downarrow} =\frac{4\pi \mathrm{M}_{s}\mathrm{t}_{\mathrm{Py}}}{\mathrm{g}\mu_{B}}(\alpha_{\mathrm{Cr_{1.12}Te_{2}/Cu/Py}}-\alpha_{\mathrm{Py}}),\label{eq4}
\end{eqnarray}
Where $4 \pi \mathrm{M}_{\mathrm{s}}$ is the saturation magnetization. The extracted values of the effective spin mixing conductance $\mathrm{g}^{\uparrow\downarrow}$ reveals the impact of interface modification using a low SOC material. Specifically, for the sample with a Cu interlayer, Py(10 nm)/Cu(2 nm)$\mathrm{/Cr_{1.12}Te_{2}}$, the value of $\mathrm{g}^{\uparrow\downarrow}$ is found to be $1.72 \times 10^{18}$ m$^{-2}$. In comparison, for the sample without the Cu interlayer, Py(10 nm)$\mathrm{/Cr_{1.12}Te_{2}}$, the value is slightly lower at  $1.67 \times 10^{18}$ m$^{-2}$. The observed marginal increase in the $\mathrm{g}^{\uparrow\downarrow}$ upon the incorporation of a 2 nm Cu interlayer at the Py(10 nm)$\mathrm{/Cr_{1.12}Te_{2}}$, interface indicates a slight enhancement in interfacial spin transfer efficiency. Despite this increase, the variation in $\mathrm{g}^{\uparrow\downarrow}$ remains minimal, suggesting that the presence of the Cu spacer layer introduces only a negligible perturbation to the overall spin transmission. The minor deviation may be attributed to interfacial spin scattering or enhanced spin resistance at the Py/Cu and Cu$\mathrm{/Cr_{1.12}Te_{2}}$ interfaces. However, the relatively close values also suggest that the Cu layer does not significantly suppress the overall spin communication between the Py and $\mathrm{Cr_{1.12}Te_{2}}$ layers.
The observed enhancement in both the $\alpha_{\mathrm{eff}}$ and the  $\mathrm{g}^{\uparrow\downarrow}$ at the Py/Cu(2 nm)$\mathrm{/Cr_{1.12}Te_{2}}$ and Py$\mathrm{/Cr_{1.12}Te_{2}}$ interface can be primarily ascribed to the generation and injection of pure spin current at room temperature from the Py layer into the adjacent  $\mathrm{Cr_{1.12}Te_{2}}$  and Cu$\mathrm{/Cr_{1.12}Te_{2}}$ interfaces, mediated via the spin pumping mechanism. At this interface, spin angular momentum is transferred from the precessing magnetization of the Py layer into the $\mathrm{Cr_{1.12}Te_{2}}$
and Cu$\mathrm{/Cr_{1.12}Te_{2}}$ interfaces, resulting in spin accumulation at the Py$\mathrm{/Cr_{1.12}Te_{2}}$ and Py/Cu$\mathrm{/Cr_{1.12}Te_{2}}$ interfaces. The accumulated spin population relaxes primarily via spin flip scattering, introducing an additional dissipation channel that enhances the spin pumping induced effective damping contribution, $\alpha_{\mathrm{sp}}$, through increased interfacial spin relaxation~\cite{Ando5.1}. The diffusive flow of spins in Py$\mathrm{/Cr_{1.12}Te_{2}}$ and Py/Cu (2 nm)$\mathrm{/Cr_{1.12}Te_{2}}$ interfaces can be described by spin current density $\mathrm{J}_{\mathrm{s}}$, evaluated using the following expression~\cite{Tserkovnyak5.1, Tserkovnyak5.3}.

{\begin{widetext}
\begin{equation}
\mathrm{J}_{\mathrm{s}} \approx \left(\frac{\mathrm{g}^{\uparrow\downarrow}\hbar}{8\pi}\right)\left(\frac{
\mathrm{h}_{\mathrm{rf}}\gamma}{\alpha}\right)^2 \left[\frac{4\pi \mathrm{M}_{\mathrm{s}}\gamma+\sqrt{(4\pi\mathrm{M}_{\mathrm{s}}\gamma)^2+16(\pi \mathrm{f})^2}}{(4\pi\mathrm{M}_{\mathrm{s}}\gamma)^2+16(\pi \mathrm{f})^2}\right]\left(\frac{2e}{\hbar}\right), \label{eq5}
 \end{equation} 
\end{widetext}}

Here $\mathrm{h}_{\mathrm{rf}}$ is the RF magnetic field of $1$ Oe (at 15 dBm rf power) in the strip line of the CPW. The calculated values of $\mathrm{J}_{\mathrm{s}}$ are found to be 0.298$\pm $0.003 MA/m$^{2}$ for  $\mathrm{Py/Cu/Cr_{1.12}Te_2}$ the interface and 0.242$\pm$0.004 MA/m$^{2}$ for $\mathrm{Py/Cr_{1.12}Te_2}$ the interface.
\begin{figure*}
\includegraphics[width=0.99\linewidth]{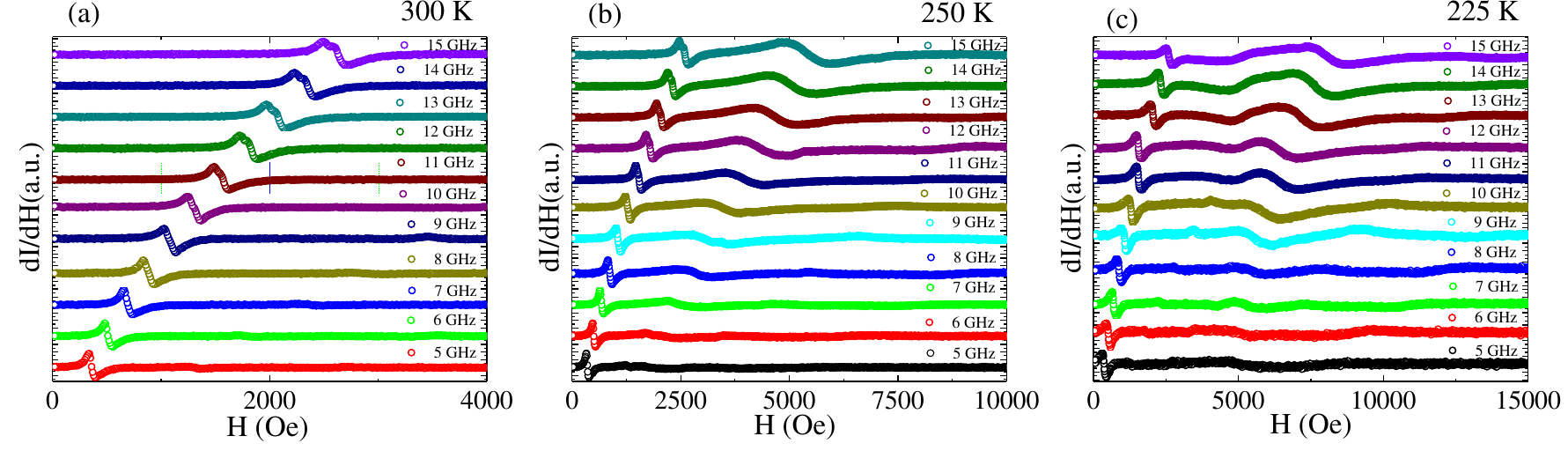}
\caption{FMR spectra were recorded for the Py/Cu/$\mathrm{Cr_{1.12}Te_2}$ sample at (a) 300 K, (b) 250 K, and (c) 225 K.}
\label{fig5}
\end{figure*}
 \begin{figure}
\includegraphics[width=0.99\linewidth]{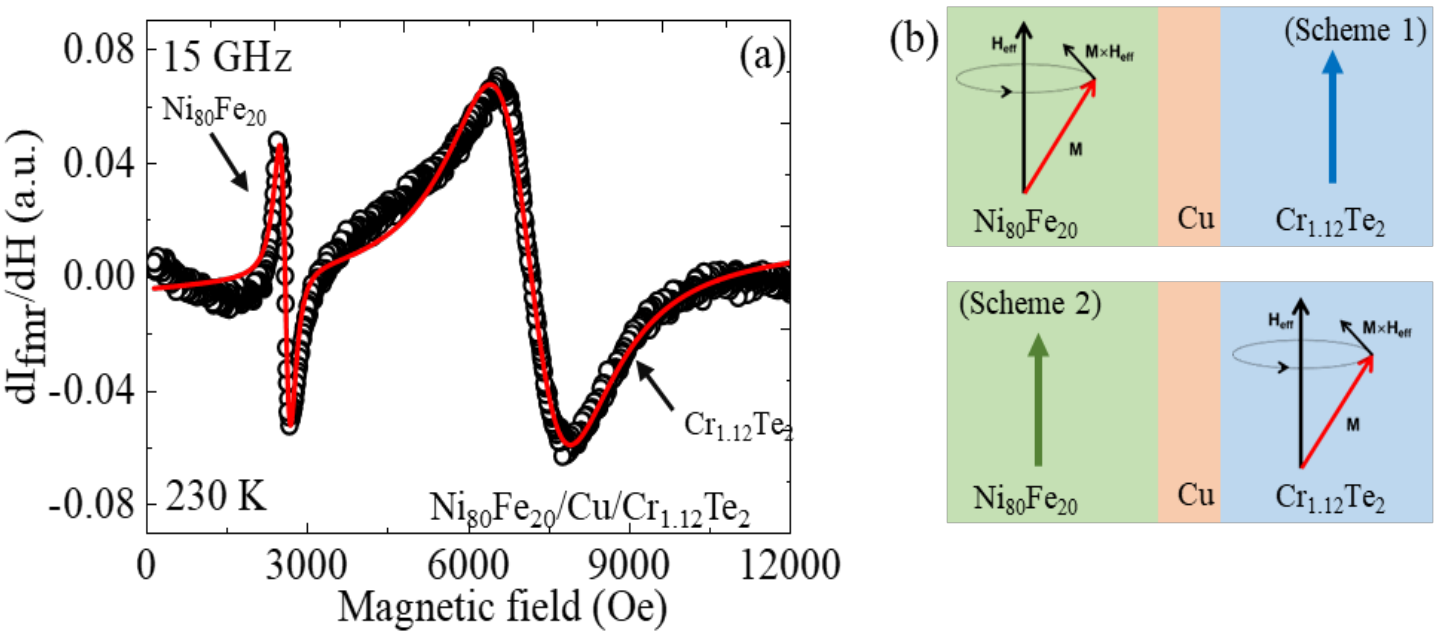}
\caption{(a) FMR spectrum of the trilayer stack $\mathrm{Ni_{80}Fe_{20}/Cu/Cr_{1.12}Te_{2}}$ measured at a frequency of 15 GHz and a temperature of 230 K. (b) Schematic illustration of the $\mathrm{Ni_{80}Fe_{20}/Cu/Cr_{1.12}Te_{2}}$ trilayer,  showing the individual magnetization precession in $\mathrm{Ni_{80}Fe_{20}}$ (Scheme 1) and $\mathrm{Cr_{1.12}Te_{2}}$ (Scheme 2), respectively.}
\label{fig6}
\end{figure}  
\section{ Temperature-dependent Spin current exchange at $\mathrm{Ni_{80}Fe_{20}/Cu/Cr_{1.12}Te_2}$ interface}
 
In a $\mathrm{FM1/NM/FM2}$ spin valve structure, the use of copper (Cu) as the nonmagnetic (NM) spacer layer ensures high interfacial spin transparency, attributed to Cu’s exceptional electrical conductivity and intrinsically weak SOC. This high transparency enables the efficient transmission of pure spin currents across the FM/NM interfaces with minimal spin memory loss, thereby preserving spin coherence during transport. The spin current generated by the precessing magnetization of FM1 is transmitted more effectively through the Cu layer. Owing to the Cu’s weak spin relaxation, spin current dissipation within the NM layer is relatively low, allowing a larger fraction of the pumped spin current to reach FM2~\cite{Tserkovnyak5.2}. The second ferromagnetic layer (FM2), acting as a spin sink, efficiently absorbs the transverse spin components, thereby enhancing the spin transfer torque (STT) exerted on its magnetization~\cite{Wei5}. This efficient absorption in FM2 leads to a notable increase in the effective Gilbert damping of FM1. Therefore, in FM1/Cu/FM2 systems, the combination of minimal spin scattering in Cu and strong spin current absorption in FM2 allows for highly efficient spin pumping, critically influencing the magnetization dynamics in such multilayered magnetic stacks~\cite{Tserkovnyak5.1}. In magnetic trilayer systems (FM1/NM/FM2), spin pumping is typically modeled as a reciprocal process, ${\mathrm{FM1}\Leftrightarrow\mathrm{FM2}}$, governed by a single, interfacial spin mixing conductance $\mathrm{g}^{\uparrow\downarrow}$ common mode for both interfaces. This approximation remains valid when FM1 and FM2 are compositionally homogeneous, resulting in structurally symmetric and magnetically equivalent FM/NM interfaces~\cite{Heinrich5,Yang5,Tserkovnyak5.2}. Many spintronic devices are based on multilayered magnetic structures, where in the FM1 and FM2 typically consist of distinct materials with differing interfaces adjacent to the NM spacer~\cite{Sani5,Mohseni5,Deac5,Heinrich5}. In such asymmetric ferromagnetic stacks, the two ferromagnetic constituents exhibit different intrinsic magnetic properties, including variations in uniaxial and cubic anisotropies, shape anisotropy, and saturation magnetization. This asymmetry necessitates treating the spin pumping mechanism as inherently nonreciprocal. In these systems, the efficiency of spin current transmission across the interfaces is governed by the spin mixing conductance $\mathrm{g}^{\uparrow\downarrow}$, which is highly sensitive to the magnetic properties of the interface. Owing to the disparities in both the magnetic and crystallographic properties of FM1 and FM2, the ($\mathrm{g}^{\uparrow\downarrow}$) acquires a directional dependence. For spin current transmission from FM1 to FM2, denoted as ($\mathrm{g}^{\uparrow\downarrow})_{\mathrm{FM1} \rightarrow \mathrm{FM2}}$, differs in magnitude and possibly phase from the conductance in the reverse direction, ($\mathrm{g}^{\uparrow\downarrow})_{\mathrm{FM2} \rightarrow \mathrm{FM1}}$, reflecting the asymmetry at the interfaces on either side of the NM spacer. This non reciprocity influences spin dynamics, effective Gilbert damping, and torque effects and must be considered in theoretical and experimental direction dependent analysis~\cite{Y5,Heinrich5}. 

We explores the temperature dependent spin current exchange at the Py/Cu$\mathrm{/Cr_{1.12}Te_2}$ (FM1/NM/FM2) interface consisting of Py, Cu, and $\mathrm{Cr_{1.12}Te_2}$, respectively. The Py and $\mathrm{Cr_{1.12}Te_2}$ layers act as dynamic sources of spin current, generated through magnetization precession especially under FMR conditions, when $\mathrm{Cr_{1.12}Te_2}$ exhibits ferromagnetic ordering. Adjacent to this is a thin Cu spacer layer, which is commonly employed in spintronic devices as NM for the transmission of spin current. Since, Cu is highly conductive and exhibits long spin diffusion lengths and negligible intrinsic SOC, making it an ideal for transmitting spin polarized electrons without introducing significant scattering or decoherence. Here, we have neglected the torque term arising from dipolar or indirect exchange interaction due to the Cu insertion layer. Ignoring the spin flip scattering probability at both Py/Cu$\mathrm{/Cr_{1.12}Te_2}$ interfaces and assuming Cu to be completely spin transparent. This system presents a compelling platform for exploring thermally modulated spin transport mechanisms due to the combination of a soft FM metal (Py), a NM spacer (Cu), and a 2D magnetic material $\mathrm{Cr_{1.12}Te_2}$.

\subsection{Temperature dependent transfer of spin angular momentum from the Py to $\mathrm{Cr_{1.12}Te_2}$}

FMR spectra of the Py/Cu$\mathrm{/Cr_{1.12}Te_2}$ interface, recorded over a broad frequency range of (5–15 GHz) at various temperatures (300 K, 250 K, and 225 K) are shown in Fig.~\ref{fig5}. The derivative Lorentzian function Eq.~\ref{eq1} having symmetric and asymmetric coefficients, is used to fit the recorded FMR spectra. Figure~\ref{fig6}(a) presents the FMR spectrum of the Py/Cu$\mathrm{/Cr_{1.12}Te_2}$ interface measured at a frequency of 15 GHz at 230 K. The spectrum exhibits two distinctly separated resonance peaks observed at $\approx$ 3000 Oe and 8000 Oe, indicating that the magnetization dynamics of the two FMs are largely decoupled. Specifically, when the magnetization vector of the Py layer is precessing at maximum amplitude (Scheme 1) of Fig~\ref{fig6}(b), the $\mathrm{Cr_{1.12}Te_2}$ layer remains nearly static, and conversely, when the magnetization of $\mathrm{Cr_{1.12}Te_2}$ layer is precessing at maximum amplitude (Scheme 2), the Py magnetization is essentially stationary, as shown in Fig~\ref{fig6}(b). The precession of FM layers adjacent to a Cu spacer is expected to induce spin accumulation at the respective FM/Cu interfaces, which can give rise to an effective magnetic field~\cite{Manchon5}. 

\begin{figure}
\includegraphics[width=0.99\linewidth]{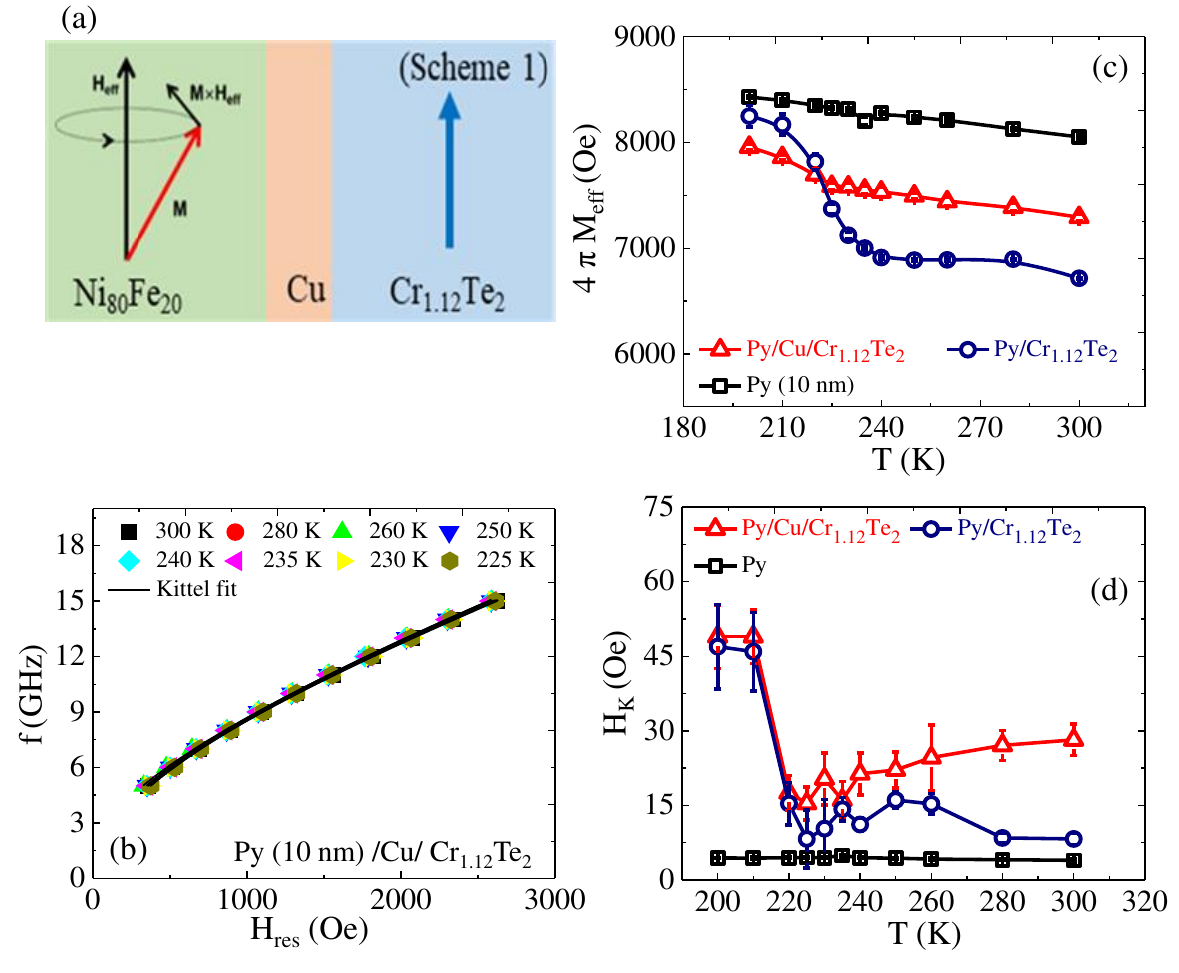}
\caption{(a)Schematic of the $\mathrm{Ni_{80}Fe_{20}/Cu/Cr_{1.12}Te_2}$ trilayer showing spin pumping and spin absorption by $\mathrm{Cr_{1.12}Te_2}$.
(b)}
\label{fig7}
\end{figure}

In the Py/Cu$\mathrm{/Cr_{1.12}Te_2}$ interface, the precessing magnetization of Py injects a pure spin current into the Cu layer. This spin current diffuses through the Cu spacer and reaches the Cu/$\mathrm{Cr_{1.12}Te_2}$ interface, where it interacts with the localized magnetic moments in $\mathrm{Cr_{1.12}Te_2}$ through interfacial exchange coupling. The efficiency of spin current transfer and absorption at the Cu/$\mathrm{Cr_{1.12}Te_2}$ interface is highly sensitive to the temperature. When the Py and $\mathrm{Cr_{1.12}Te_2}$ precess at their individual $\mathrm{H_{res}}$, spin accumulation occurs at the Py/Cu and $\mathrm{Cr_{1.12}Te_2/Cu}$ interfaces, respectively. However, if both FMs are simultaneously precess at the same $\mathrm{H_{res}}$, the spin currents traversing the Py/Cu and $\mathrm{Cr_{1.12}Te_2/Cu}$ interfaces cancels each other, resulting in no spin accumulation at the interfaces. To quantitatively analyze the resonance behavior of each layer, the resonance fields $\mathrm{H_{res}}$ and linewidth $\Delta \mathrm{H}$ were determined by de-convoluting the composite FMR signal into two distinct complex components. This was achieved by fitting the derivative of the sum of symmetric and antisymmetric Lorentzian function, as described by Eq.~\ref{eq1}. Figure~\ref{fig7} (a) shows the transfer of spin angular momentum from the precessing magnetization of the Py layer to the Cu$\mathrm{/Cr_{1.12}Te_2}$ interface. Figure~\ref{fig7}(b) presents the dependence of the $\mathrm{H}_{\mathrm{res}}$, on $\mathrm{f}$, measured across a broad temperature range from 225 K to 300 K. The effective magnetization, $4\pi \mathrm{M}_{\mathrm{eff}}$ and $\mathrm{H}_{\mathrm{K}}$, of the Py layer at various temperatures are quantitatively extracted by fitting the experimental data to the Kittel Eq.~\ref{eq3}. The variation of $4\pi\mathrm{M}_{\mathrm{eff}}$ with temperature for the Py, Py$\mathrm{/Cr_{1.12}Te_2}$ and Py/Cu$\mathrm{/Cr_{1.12}Te_2}$ samples is shown in Fig.~\ref{fig7}(c). It is observed that the values of  $4\pi \mathrm{M}_{\mathrm{eff}}$ increases as the temperature decreases. This trend can be attributed to the temperature dependence of the saturation magnetization $4\pi \mathrm{M}_{\mathrm{s}}$ of the Py layer. As the temperature decreases, thermal agitation of magnetic moments is reduced, leading to stronger alignment of the spins. The $4\pi \mathrm{M}_{\mathrm{s}}$ increases, which directly contributes to an increase in the $4\pi \mathrm{M}_{\mathrm{eff}}$ values. In samples such as Py/Cr$_{1.12}$Te$_2$ and Py/Cu/Cr$_{1.12}$Te$_2$, interfacial exchange coupling at low temperatures may contribute to the observed behavior. A sudden enhancement of $4\pi \mathrm{M}_{\mathrm{eff}}$ values near $\mathrm{T_{c}}$ is observed for Py/Cr$_{1.12}$Te$_2$ which suggests interfacial exchange coupling. In the Py/Cu$\mathrm{/Cr_{1.12}Te_2}$  samples, the presence of the  Cu spacer layer  decouples the Py and $\mathrm{Cr_{1.12}Te_2}$ layers, thereby suppressing direct interfacial exchange interactions due to the lack of direct magnetic contact. These interactions are more pronounced in the Py$\mathrm{/Cr_{1.12}Te_2}$ samples, where the Py and $\mathrm{Cr_{1.12}Te_2}$ layers are in direct contact.

The strength of such interfacial exchange interactions typically increases at lower temperatures, potentially reinforcing magnetic order and leading to an increase in $4\pi \mathrm{M}_{\mathrm{eff}}$. This effect can be particularly pronounced in structures incorporating magnetic and correlated materials like Cr$_{1.12}$Te$_2$, which exhibit magnetic transitions and enhanced spin correlations in the low temperature regimes. Figure~\ref{fig7} (d)  shows the temperature dependence of the $\mathrm{H}_{\mathrm{K}}$ for Py, Py$\mathrm{/Cr_{1.12}Te_2}$ and Py/Cu$\mathrm{/Cr_{1.12}Te_2}$ samples. In the case of ref. Py, $\mathrm{H}_{\mathrm{K}}$ remains the same across the temperature range, attributed to its intrinsically low magneto crystalline anisotropy and minimal thermal variation in magnetic properties. In contrast, both Py$\mathrm{/Cr_{1.12}Te_2}$ and Py/Cu$\mathrm{/Cr_{1.12}Te_2}$ interfaces exhibit a temperature dependence of   $\mathrm{H}_{\mathrm{K}}$, which could be due to interfacial exchange coupling with the $\mathrm{Cr_{1.12}Te_2}$ layer. The modulation of $\mathrm{H}_{\mathrm{K}}$ vs. temperature in these samples shows the critical role of interfacial magnetic effects and thermally driven spin-dependent interactions.

\begin{figure}
\includegraphics[width=0.99\linewidth]{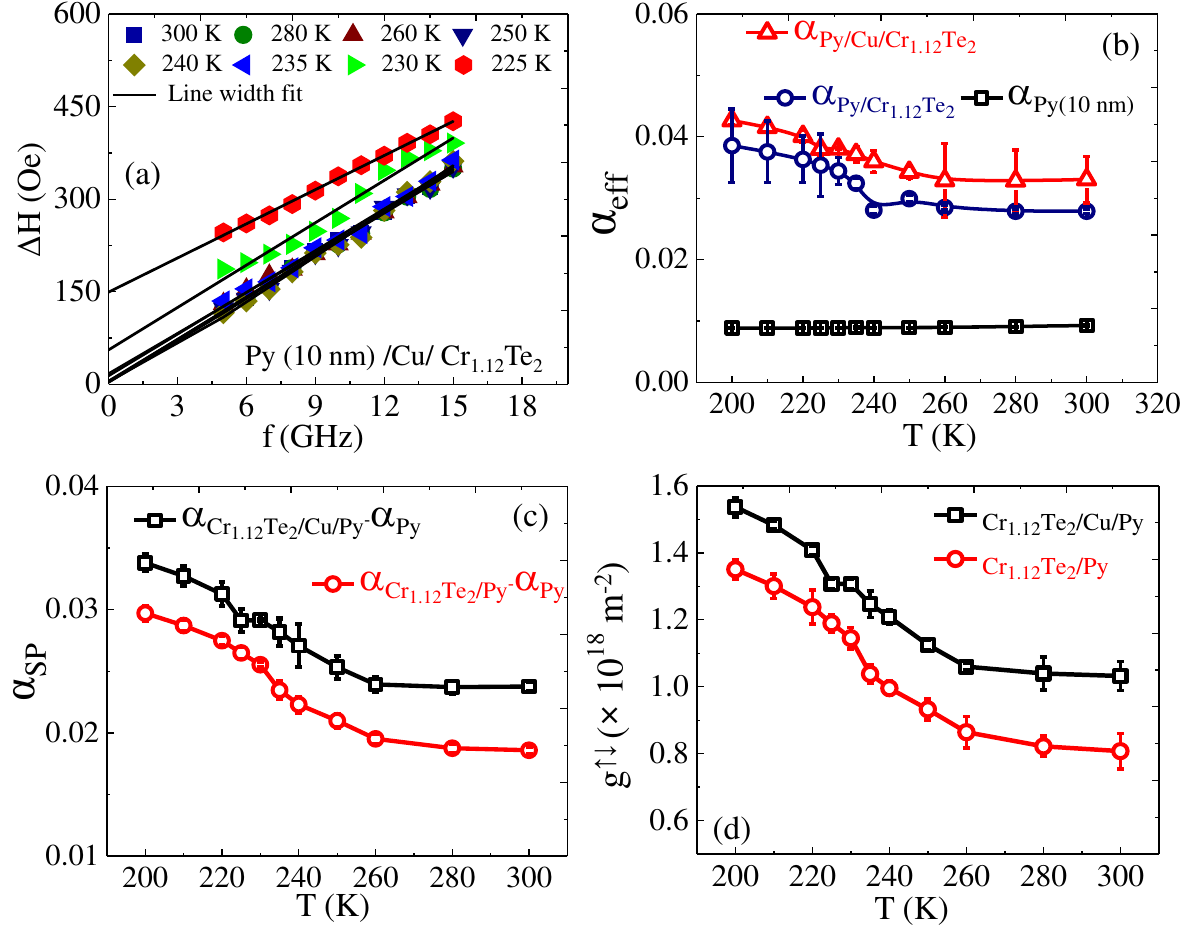}
\caption{(a) FMR spectrum of the trilayer stack $\mathrm{Ni_{80}Fe_{20}/Cu/Cr_{1.12}Te_{2}}$ measured at a frequency of 15 GHz and a temperature of 230 K. (b) Schematic illustration of the $\mathrm{Ni_{80}Fe_{20}/Cu/Cr_{1.12}Te_{2}}$ trilayer,  showing the individual magnetization precession in $\mathrm{Ni_{80}Fe_{20}}$ (Scheme 1) and $\mathrm{Cr_{1.12}Te_{2}}$ (Scheme 2), respectively.}
\label{fig8}
\end{figure}

Figure.~\ref{fig8} (a), shows the variation of $\Delta\mathrm{H}$ vs. f for Py/Cu$\mathrm{/Cr_{1.12}Te_{2}}$ samples across a broad temperature range (225-300 K). The solid lines represent linear fits based on Eq.~\ref{eq2}, which is used to extract the effective Gilbert damping factor $\mathrm{\alpha_{eff}}$. The extracted values of $\mathrm{\alpha_{eff}}$ with respect to the temperature for three different sample ref. Py, Py/$\mathrm{Cr_{1.12}Te_2}$ and Py/Cu$\mathrm{/Cr_{1.12}Te_2}$ are shown in Figure.~\ref{fig8} (b).  the case of the ref. Py sample, the damping factor remains essentially constant with temperature, indicating stable intrinsic damping properties. However, for both Py/$\mathrm{/Cr_{1.12}Te_2}$ and Py/Cu$\mathrm{/Cr_{1.12}Te_2}$ samples, a significant enhancement in $\mathrm{\alpha_{eff}}$ is observed as the temperature approaches the Curie temperature $\mathrm{T_{c}}$ of the $\mathrm{Cr_{1.12}Te_{2}}$ layer. This sharp enhancement in damping is attributed to increased spin pumping or enhanced spin scattering at the interface due to magnetic fluctuations near $\mathrm{T_{c}}$ . The presence of the Cu interlayer in Py/Cu/$\mathrm{Cr_{1.12}Te_{2}}$ slightly modifies this behavior, possibly by influencing the interfacial spin transparency.

The efficiency of spin current transfer and absorption at the Cu/$\mathrm{Cr_{1.12}Te_2}$ interface exhibits a pronounced dependence on temperature, primarily due to the intrinsic magnetic phase behavior of the $\mathrm{Cr_{1.12}Te_2}$ layer. As a 2D vdW FM, $\mathrm{Cr_{1.12}Te_2}$ undergoes a magnetic phase transition near its ($T_c$), where its long-range FM order gradually diminishes with increasing temperature. This magnetic ordering plays a crucial role in the spin transport dynamics at the interface. When a spin current is injected int o the Cu/$\mathrm{Cr_{1.12}Te_2}$  interface typically via spin pumping from an adjacent Py layer the interfacial spin mixing conductance ($\mathrm{g}^{\uparrow\downarrow}$) determines how efficiently the angular momentum of the spin current is transferred into the magnetic layer. The value of ($\mathrm{g}^{\uparrow\downarrow}$) is strongly influenced by the spin susceptibility and magnetic coherence of the sink layer. At low temperatures, where $\mathrm{Cr_{1.12}Te_2}$ is FM ordered, the density of available spin states and the alignment of magnetic moments promote strong spin exchange interactions at the interface. This enhances $\mathrm{g}^{\uparrow\downarrow}$ and makes the $\mathrm{Cr_{1.12}Te_2}$ layer an efficient spin sink, capable of absorbing and dissipating spin current. The temperature reaches $T_c$, the magnetic order in $\mathrm{Cr_{1.12}Te_2}$  weakens due to increased thermal fluctuations, reducing the availability of aligned spins for exchange interactions. As a result, the spin mixing conductance decreases, and the interfacial absorption of spin current is suppressed. In this paramagnetic or weakly ordered regime, the interface becomes more reflective to spin current, diminishing the damping enhancement in the adjacent FM layer. 

\subsection{Temperature dependent spin pumping from the $\mathrm{Cr_{1.12}Te_2}$ to $\mathrm{Ni_{80}Fe_{20}}$}

\section{Conclusion}
The $\mathrm{Ni_{80}Fe_{20}/Cu/Cr_{1.12}Te_2}$ trilayer system presents a promising platform for exploring temperature dependent spin current dynamics at hybrid interfaces. By coupling a conventional metallic FM with a 2D magnetic layer through a  Cu spacer, this structure enables precise control over spin current absorption via thermal modulation of magnetic order in $\mathrm{Cr_{1.12}Te_2}$. The temperature-sensitive spin mixing conductance at the Py/Cu$\mathrm{/Cr_{1.12}Te_2}$ and Py$\mathrm{/Cr_{1.12}Te_2}$ interface directly influences the effective damping in the FM layer, offering a tunable mechanism for manipulating spin transport. Moreover, the atomically smooth and chemically stable nature of the 2D $\mathrm{Cr_{1.12}Te_2}$ ensures coherent spin transmission and minimal interfacial scattering. These features make the trilayer architecture not only a valuable system for fundamental spintronic studies but also a potential building block for thermally adaptive, low-power spin-based devices.

\section{Acknowledgements}
M.T. acknowledges the MHRD, Government of India for Senior Research Fellowship. S.M. acknowledges the Department of Science and Technology (DST) Nanomission, Government of India for financial support. R.M. acknowledges Initiation grant, IIT Kanpur (IITK/PHY/2022027), and I-HUB Quantum Technology Foundation (I-HUB/PHY/2023288), IISER Pune  for financial support.

\end{document}